\documentclass[12pt,preprint]{aastex}
\begin{document}

\newcommand{\kms}{\ensuremath{\mathrm{km}\,\mathrm{s}^{-1}}}
\newcommand{\etal}{et al.}
\newcommand{\THmod}{Tuorla-Heidelberg}
\newcommand{\LCDM}{$\Lambda$CDM}
\newcommand{\ML}{\ensuremath{\Upsilon_{\star}}}
\newcommand{\MLmax}{\ensuremath{\Upsilon_{max}}}
\newcommand{\MLpop}{\ensuremath{\Upsilon_{pop}}}
\newcommand{\MLopt}{\ensuremath{\Upsilon_{acc}}}
\newcommand{\Om}{\ensuremath{\Omega_m}}
\newcommand{\Ob}{\ensuremath{\Omega_b}}
\newcommand{\OL}{\ensuremath{\Omega_{\Lambda}}}
\newcommand{\C}{\ensuremath{\mathfrak{C}}}
\newcommand{\B}{\ensuremath{\mathfrak{B}}}
\newcommand{\Q}{\ensuremath{{\cal Q}}}
\newcommand{\Pop}{\ensuremath{{\cal P}}}
\newcommand{\G}{\ensuremath{{\Gamma}}}
\newcommand{\Lsun}{\ensuremath{L_{\odot}}}
\newcommand{\rd}{\ensuremath{R_{d}}}
\newcommand{\Msun}{\ensuremath{{\cal M}_{\odot}}}
\newcommand{\mass}{\ensuremath{{\cal M}}}
\newcommand{\Mst}{\ensuremath{{\cal M}_{\star}}}
\newcommand{\Mg}{\ensuremath{{\cal M}_g}}
\newcommand{\Md}{\ensuremath{{\cal M}_d}}
\newcommand{\Mh}{\ensuremath{{\cal M}_h}}
\newcommand{\Sd}{\ensuremath{{\Sigma}_0}}
\newcommand{\Sg}{\ensuremath{{\Sigma}_g}}
\newcommand{\Sst}{\ensuremath{{\Sigma}_{\star}}}
\newcommand{\fst}{\ensuremath{f_{\star}}}
\newcommand{\Vst}{\ensuremath{V_{\star}}}
\newcommand{\vst}{\ensuremath{v_{\star}}}
\newcommand{\Vg}{\ensuremath{V_{g}}}
\newcommand{\vt}{\ensuremath{v_{t}}}
\newcommand{\Vc}{\ensuremath{V_{c}}}
\newcommand{\Vb}{\ensuremath{V_{b}}}
\newcommand{\Vh}{\ensuremath{V_h}}
\newcommand{\Vf}{\ensuremath{V_{f}}}
\newcommand{\VNFW}{\ensuremath{V_{200}}}
\newcommand{\shape}{\ensuremath{\mathfrak{T}_{0.6}}}
\newcommand{\norm}{\ensuremath{\sigma_8}}
\newcommand{\gn}{\ensuremath{g_N}}
\newcommand{\anot}{\ensuremath{a_0}}
\newcommand{\csb}{\ensuremath{\mu_0}}
\newcommand{\magsq}{\ensuremath{\mathrm{mag.}\,\mathrm{arcsec}^{-2}}}
\newcommand{\surfdens}{\ensuremath{{\cal M}_{\odot}\,\mathrm{pc}^{-2}}}
\newcommand{\MDAC}{MDAcc}
\newcommand{\HI}{H{\sc i}}


\title{Milky Way Mass Models and MOND}

\author{Stacy S.~McGaugh} 

\affil{Department of Astronomy, University of Maryland}
\affil{College Park, MD 20742-2421}    
\email{ssm@astro.umd.edu}

\begin{abstract}
Using the \THmod\ model for the mass distribution of the Milky Way,
I determine the rotation curve predicted by MOND.  The result is in good
agreement with the observed terminal velocities interior to the solar radius
and with estimates of the Galaxy's rotation curve exterior thereto.  
There are no fit parameters: given the mass distribution, MOND provides
a good match to the rotation curve.  The \THmod\ model 
does allow for a variety of exponential scale lengths;  MOND 
prefers short scale lengths in the range $2.0 \lesssim \rd \lesssim 2.5$ kpc.  
The favored value of \rd\ depends somewhat on the choice of interpolation
function.  There is some preference for the `simple' interpolation function
as found by Famaey \& Binney.  I introduce an interpolation
function that shares the advantages of the simple function on galaxy scales
while having a much smaller impact in the solar system.  I also solve the
inverse problem, inferring the surface mass density distribution of the Milky Way
from the terminal velocities.  
The result is a Galaxy
with `bumps and wiggles' in both its luminosity profile and rotation curve
that are reminiscent of those frequently observed in external galaxies.
\end{abstract}


{~}

\clearpage

\section{Introduction}

A persistently viable alternative to dark matter is the Modified Newtonian
Dynamics (MOND) of Milgrom (1983abc).  MOND is known to perform well
in fitting the rotation curves of individual galaxies 
(e.g., Begeman, Broeils, \& Sanders 1991; Sanders 1996; 
de Blok \& McGaugh 1998; Sanders \& Verheijen 1998;
Sanders \& McGaugh 2002; Sanders \& Noordermeer 2007), 
possessing a predictive
power well beyond that available from mass models with a mix of luminous
and dark mass.  On the other hand, MOND fares less well on the larger scales
of clusters of galaxies (e.g., Sanders 2003; Pointecouteau \& Silk 2005;
Clowe \etal\ 2006; Angus, Famaey, \& Buote 2007),
seeming to require more mass than currently observed in known baryonic forms.
This is a genuine problem, but also a partial success as Milgrom (1983c) was
among the first to note the potential mass of the intracluster medium.  
Clusters are not without their puzzles in the context of dark matter
(Hayashi \& White 2006; McCarthy, Bower, \& Balogh 2007; 
Springel \& Farrar 2007; Milosavljevi{\'c} \etal\ 2007; Milgrom \& Sanders 2007; 
Angus \& McGaugh 2008; Nusser 2008; Broadhurst \& Barkana 2008).
This is an oft-repeated theme;
systems that pose problems for MOND often make little sense in terms of
dark matter either.  On the positive side, 
MOND performs well in a greater variety of systems 
than seems to be widely appreciated 
(Bekenstein 2006; McGaugh 2006; Milgrom 2008).  For example,
it correctly describes (Milgrom 2007; Gentile \etal\ 2007) the behavior
of tidal dwarfs (Bournaud \etal\ 2007) that pose an existential challenge
to cold dark matter.  We should therefore be cautious of an
eagerness to dismiss the idea entirely on the basis of one type of object, 
or the natural distrust of the unfamiliar, and 
rather seek as many precision tests as possible.

As a modified force law, MOND makes strong and testable predictions.
In the absence of invisible mass, the observed kinematics of an object in the
MOND regime of low acceleration should follow from its observed mass 
distribution just as planetary motions follow from purely Newton dynamics
in the  solar system.  One galaxy that we are intimately familiar with is
our own Milky Way, for which a wealth of high quality data are available.

Famaey \& Binney (2005) investigated MOND in the Milky Way in the
context of the Basel model (Bissantz \& Gerhard 2002;
Bissantz, Englmaier \& Gerhard 2003).
Since that time, the number and quality of observational constraints has
continued to improve.  Moreover, the gas mass is not negligible in MOND,
being nearly 20\% of the total.  Here I consider this in the context
of the Milky Way for the first time.  

The goal of this paper is to use one type of observation to predict another.
Given the mass distribution of the Milky Way, does MOND predict a plausible
rotation curve consistent with the available data?  Inverting the question, can
the mass distribution be inferred from the kinematic data?  The point is not to
find the best fit involving every possible kind of data (see Widrow, Pym,
\& Dubinski 2008 for
such an exercise in the conventional context).  Rather, I seek to use one type of
data to \textit{predict} an entirely independent set of observations.

The paper is organized as follows.
In section \S \ref{mus}, I describe MOND as applied here, considering a variety
of interpolation functions.
In \S \ref{MWmass} I describe the mass distribution of the Milky Way,
adopting the \THmod\ model (Flynn \etal\ 2006) for the stellar mass distribution,
and utilizing the tabulation of Olling \& Merrifield (1999) for the gas distribution.
In \S \ref{MWRC} I present the results of applying MOND to the adopted mass
model.  In \S \ref{MW_emp} I invert the question, and infer the deviations from
a smooth exponential luminosity profile implied by the bumps and wiggles in the
terminal velocity curve.  The conclusions are summarized in \S \ref{conc}.

\section{MOND and the Interpolation Function \label{mus}}

The basic idea of MOND (Milgrom 1983a) is that rather than invoking dark
matter to explain mass discrepancies in extragalactic systems, one modifies
the force law such that
\begin{equation}
\textbf{g}_N = \mu(x) \textbf{a}, \label{mondeqn}
\end{equation}
where \textbf{g}$_N$ is the acceleration calculated using Newtonian dynamics,
and $\textbf{a}$ is the actual resultant acceleration.
The modification occurs not at some length scale, but at the acceleration 
scale \anot, with $x = a/\anot$ (here $a = |\textbf{a}|$).
The interpolation function $\mu(x)$ has the property $\mu \rightarrow 1$
in the limit of large accelerations $a \gg \anot$ so that Newtonian behavior
is recovered.  In the limit of small accelerations the modification applies, with
$\mu \rightarrow x$ for $a \ll \anot$.  

The acceleration scale \anot\ is very small.
Begeman \etal\ (1991) found $\anot = 1.2 \times 10^{-10}\;
\textrm{m}\,\textrm{s}^{-2} = 
3700\;\textrm{km}^2\,\textrm{s}^{-2}\,\textrm{kpc}^{-1}$ from fits to high
quality rotation curves.  More recently, Bottema \etal\ (2002) estimated
$\anot = 3000\;\textrm{km}^2\,\textrm{s}^{-2}\,\textrm{kpc}^{-1}$
and McGaugh (2004) 
$\anot \approx 4000\;\textrm{km}^2\,\textrm{s}^{-2}\,\textrm{kpc}^{-1}$,
the latter utilizing population synthesis estimates of the stellar mass.  Here I
adopt the intermediate Begeman \etal\ (1991) value and keep \anot\ fixed
while considering different possible interpolation functions.  The reader should
bear in mind that in addition to the uncertainty in \anot\ reflected
in the variety of determinations, the best fit value is likely to depend somewhat
on the choice of interpolation function.

As noted by Milgrom (1983a) and Felten (1984),
equation~\ref{mondeqn} does not conserve momentum.  
This was addressed by Bekenstein \& Milgrom (1984), who
wrote the modified Poisson equation
\begin{equation}
\nabla\cdot\left[\mu\left(\frac{|\nabla\Phi|}{\anot}\right)\nabla\Phi\right]  = 4 \pi G \rho.  
\label{AQUAL}  
\end{equation}
This form of modified gravity obeys the conservation laws.
Milgrom (1994, 1999) also provides a conservative albeit non-local formalism
for modified inertia rather than modified gravity.  

Application of  equation~\ref{mondeqn} has been highly 
successful in fitting rotation curves.
It is exact for circular orbits in the modified inertia theory.
For modified gravity as it applies in spiral galaxies, 
equation~\ref{mondeqn} is an approximation to equation~\ref{AQUAL}
that is usually correct to $\sim 15\%$ (Brada \& Milgrom 1995).
For our considerations here this suffices; it is not necessary to invoke the 
aquadratic Lagrangian theory of Bekenstein \& Milgrom (1984),
much less the generally covariant theories of Bekenstein (2004) or Sanders (2005).

Indeed, for our purposes here, we merely need the empirically proven formula
connecting surface density and rotation velocity (McGaugh 2004).  MOND is
formulated in terms of the actual acceleration, as appropriate for a dynamical
theory.  However, in the Milky Way we have
a better handle on the surface densities that predict the Newtonian acceleration.
Therefore, it is convenient to make the substitution $\nu(y) = \mu^{-1}(x)$, where
$y = \gn/\anot$.  While not appropriate as the basis for a theory, this is
functionally equivalent when using the empirical approximation of
equation~\ref{mondeqn}.  Replacing $\mu(x)$ with $\nu^{-1}(y)$ has the 
advantage that $\gn(R)$ can be directly computed
from $\Sigma(R)$ with purely Newtonian dynamics.  
Assuming circular motion ($a = \Vc^2/R$), we then have
\begin{equation}
\Vc^2 = \nu(y) \Vb^2,	\label{nunotmu}
\end{equation}
where \Vb\ is the Newtonian rotation velocity expected for the baryons.
The Newtonian velocity $\Vb$ and acceleration \gn\ are computed for 
appropriate mass distributions (see \S \ref{MWmass}): no simplifying
assumption like a `spherical disk' is made.  Moreover, the right hand side of
equation~\ref{nunotmu} depends only on surface density, making it possible
to predict the rotation curve without reference to it through 
$x = \Vc^2/(\anot R)$.

Rotation curve fitting has traditionally employed the interpolation function
\begin{equation}
\mu_2(x) = \frac{x}{\sqrt{1+x^2}} \label{stdfcn}
\end{equation}
(Milgrom 1983b; Sanders \& McGaugh 2002).
This form is not specified by any deeper theory, and other forms are possible.
From habitual use equation~\ref{stdfcn} has come to be called the `standard'
interpolation function.

More recently, it has been suggested that the `simple' function
\begin{equation}
\mu_1(x) = \frac{x}{1+x} \label{simplefcn}
\end{equation}
may give a better description of some data (Famaey \& Binney 2005;
Zhao \& Famaey 2006; Sanders \& Norrdermeer 2007).
Both this and the standard function are part of the family
\begin{equation}
\mu_n(x) = x(1+x^n)^{-1/n}.
\end{equation}
If we make the transformation $\mu \rightarrow \nu^{-1}$, we find the 
corresponding family
\begin{equation}
\nu_n(y) = \left(\onehalf+\onehalf \sqrt{1+4y^{-n}}\right)^{1/n}.
\end{equation}

Milgrom \& Sanders (2007) have suggested other possible families.
One is
\begin{equation}
\tilde \nu_n(y) = \frac{1}{\sqrt{1-e^{-y}}} +n e^{-y},
\end{equation}
with another being
\begin{equation}
\bar \nu_n(y) = (1-e^{-y^n})^{-\frac{1}{2n}} 
	+ (1- \frac{1}{2n}) e^{-y^n}.
\end{equation}
These forms may have some virtue in causing artifacts that appear
as rings of dark matter (Milgrom \& Sanders 2007) such as that described by
Jee \etal\ (2007).  Note that $\nu(y)$ is effectively the same as the mass 
discrepancy $D(y)$ defined by McGaugh (2004), but I make the distinction
that $D$ is an empirical quantity while $\nu$ is a construct of MOND.

The simple function (equation~\ref{simplefcn}) appears to have some
important virtues in the transition from Newtonian to MOND regimes 
(Famaey \& Binney 2005; Sanders \& Noordermeer 2007).  
However, it may have difficulties with solar system tests 
(Sereno \& Jetzer 2006; Wallin, Dixon, \& Page 2007; Iorio 2007).  
This occurs because of a rather gradual approach to the Newtonian
regime.  Interestingly, it is just right (Milgrom 2006; McCulloch 2007)
 for explaining the Pioneer anomaly (Anderson \etal\ 1989).
Note that this implies a conflict in the data.  The Pioneer anomaly 
is quite accurately measured, but may not be a dynamical effect, 
so it remains unclear which interpretation to prefer.

\begin{figure}  
\epsscale{1.0}
\plotone{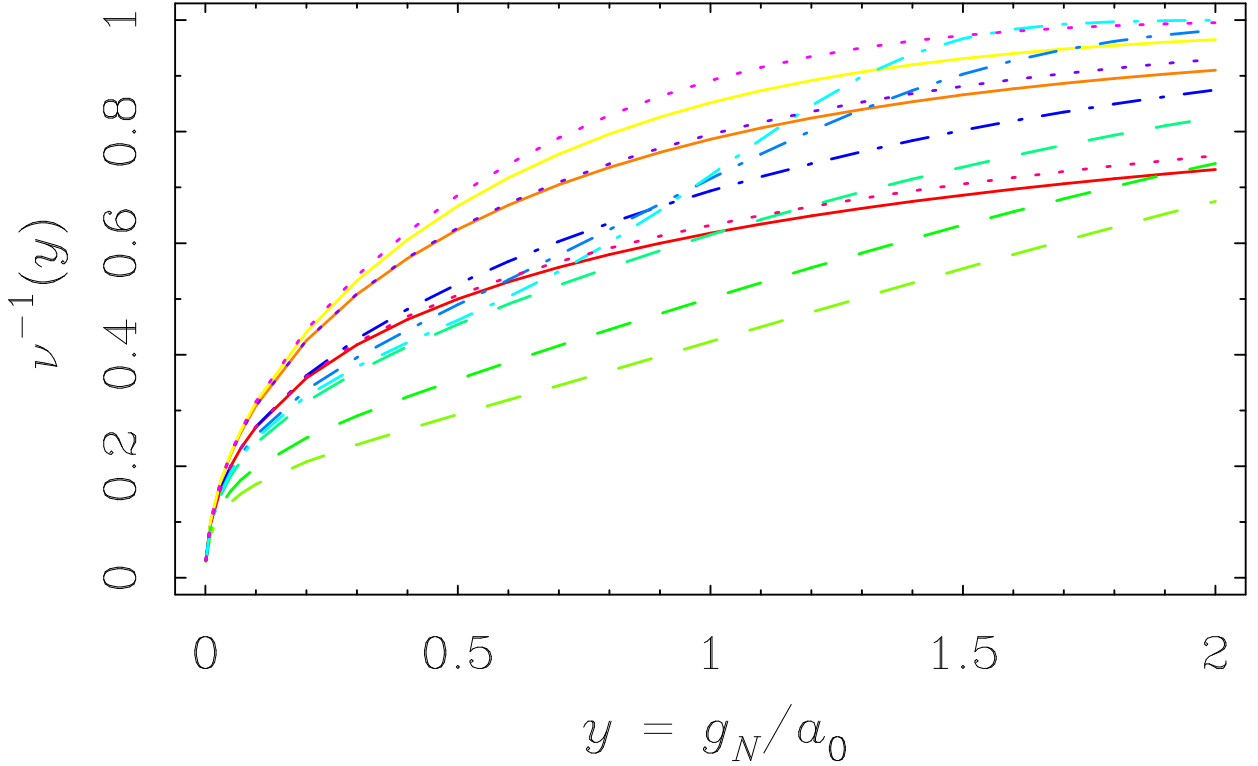}
\caption{MOND interpolation functions with the property $\nu \rightarrow 1$
for $y \gg 1$ and $\nu^{-1} \rightarrow \sqrt{y}$ for $y \ll 1$.  The traditional family
$\nu_n$ is shown as solid lines with $n$ increasing from bottom to top.
The families $\tilde \nu_n$ and $\bar \nu_n$ are shown as dashed and
dash-dotted lines, respectively.  For $\tilde \nu_n$, increasing $n$ leads to
more gradual transitions (lower lines for higher $n$, opposite the case of
the other families).  For $\bar \nu_n$,
the lines cross just shy of $y = 1$, with larger $n$ being the higher lines
at $y > 1$.  The family $\hat \nu_n$ is shown as dotted
lines, with increasing amplitude for increasing $n$.  This family is very
similar to the traditional family with the same index in the vicinity of
$y \approx 1$ relevant to galaxy dynamics, but converges to Newtonian
behavior much more rapidly at the large accelerations relevant to 
solar system dynamics.
\label{nuy}}
\end{figure}

It is possible to have an interpolation function that retains the virtues of the
simple function on galaxy scales ($\sim \anot$) while having no impact in
the inner solar system ($\sim 10^8 \anot$).  One possible family is
\begin{equation}
\hat \nu_n(y) = (1-e^{-y^{n/2}})^{-1/n}   \label{mynu}
\end{equation}
such that
\begin{mathletters}
\begin{eqnarray}
\hat \nu_1(y) = (1 - e^{-\sqrt{y}})^{-1} \label{munuone} \\
\hat \nu_2(y) = (1-e^{-y})^{-1/2}.  \label{mynutwo}
\end{eqnarray}
\end{mathletters}
On galaxy scales, $\hat \nu_1(y)$ corresponds closely to the simple function
$\nu_1(y)$ and $\hat \nu_2(y)$ to the standard function $\nu_2(y)$ 
(Fig~\ref{nuy}).  On solar system scales, the simple function predicts deviations
from purely Newtonian behavior of one part in $10^8$, while $\hat \nu_1(10^8)$
deviates by only one part in $e^{10^8}$.  While this is certainly a virtue if we
wish to avoid even modest MOND effects in the solar system, one theoretically
unpleasant aspect of $\hat \nu$ is that its transformation to $\hat \mu$ is
transcendental.  

\placefigure{nuy}

The interpolation functions given above are shown in Fig.~\ref{nuy}.
While the shapes of $\nu_n$ and $\hat \nu_n$ are similar for similar $n$,
$\tilde \nu_n$ and $\bar \nu_n$ are\footnote{There is some overlap between
families. For example, $\hat \nu_1 = \bar \nu_{1/2}$ but $\hat \nu_2 \ne
\bar \nu_{3/2}$.} rather different.  In particular, they have
a rather linear region around $y \approx 1$.  What effect this might have on
rotation curve fits has not yet been explored in detail, but it may be useful
in some historically difficult cases (Milgrom \& Sanders 2007).  The chief effect of
increasing $n$ within a given family seems to be to increase the value of
the best fit mass-to-light ratio.  This occurs because there is a more pronounced
MOND effect already at $y=1$ with $\nu_1$ than with $\nu_2$ (for example).
Thus less mass is required to attain the same velocity with lower $n$.

I do not seek here to identify the optimal version of the many possible interpolation 
functions.  Rather, I will simply illustrate the effects of a few representative
examples.  The results are grossly similar, varying only in details as 
plausible interpolation functions differ only a little.

\section{The Milky Way Mass Distribution \label{MWmass}}

To describe the stellar mass distribution of the Milky Way, I adopt the results of the
recent \THmod\ study (Flynn \etal\ 2006).  This provides a relation between the
bulge mass, disk mass, and scale length of the Milky Way (their Fig.~15)
that satisfies both the local surface density and the global luminosity.  As the
assumed scale length of the disk increases, its central surface density decreases,
and the bulge mass increases to maintain the necessary central mass 
(Table~\ref{MW_mods}).

Following Flynn \etal\ (2006), I adopt a solar radius $R_0 = 8$ kpc.
For the circular velocity I take $\Theta_0 = 219\;\kms$ (Reid \etal\ 1999).
The latter is only for reference and does not enter into the models constructed here. 
The choice of $R_0$ is relevant as it affects the total scale of the problem.
I explore a range of scale lengths \rd\ at fixed $R_0$.  
The details will change for different choices of $R_0$ since MOND imposes 
a specific physical scale, and the total mass will scale with $R_0$.
However, the differences over the plausible range of $R_0$ are not likely to
be great compared to the variation due to $\rd/R_0$. 

\placetable{MW_mods}

The particular model adopted for the baryonic mass distribution of the Milky Way
certainly matters.  For the appropriate choice of scale length, the \THmod\ model
is very similar to the Basel model (Bissantz \etal\ 2003) 
employed by Famaey \& Binney (2005).
However, even small differences are perceptible in MOND.  I employ the 
\THmod\ model here because of its novelty and 
to enable an exploration of the effects of the scale length.
The more important advance however is likely the inclusion of the gas distribution
(\S \ref{thegas}).

\subsection{The Stellar Disk \label{thedisk}}

The stellar disk is assumed to follow an exponential distribution
\begin{equation}
\Sst(R) = \Sd e^{-R/\rd}.
\end{equation}
The central surface density \Sd\ is computed from the disk mass of the
\THmod\ model for the appropriate choice of scale length (Table~\ref{MW_mods}).
The disk is assumed to have a finite thickness with an exponential vertical 
distribution with $z_d = 300$ pc (Siegel \etal\ 2002).  
The distinction between thick and thin disk is 
relatively unimportant here as most of the mass is in the thin disk.
The choice of $z_d$ is relevant only to the detailed computation of \Vb;
plausible deviations are smaller than the variation in interpolation functions. 

I consider scale lengths in the range $2 \leq \rd \leq 4$ kpc.
At the lower limit of this range, the \THmod\ model has a purely exponential
surface density distribution.  At the upper limit of this
range, the Milky Way disk starts to become rather low surface brightness.
The scale length suggested by COBE $L$-band data is 2.5 kpc 
(Binney, Gerhard, \& Spergel 1997),
while more recently Gerhard (2002, 2006) has suggested a shorter $\rd = 2.1$ kpc.
In varying the scale length, it turns out that MOND prefers a rather short scale length
consistent with these estimates, the precise value depending somewhat on the 
choice of interpolation function and the choice of comparison data.

\subsection{The Bulge \label{thebulge}}

The distribution of bulge mass follow the triaxial distribution determined
by Binney \etal\ (1997): 
\begin{equation}
\rho_B(b) = \frac{\rho_{B,0}}{\eta \zeta b_m^3}
\frac{e^{-(b/b_m)^2}}{\left(1+b/b_0\right)^{1.8}}
\label{bulgedist}
\end{equation}
where $b^2 = x^2+(y/\eta)^2+(z/\zeta)^2$.  They base this profile
on fits to the COBE $L$-band data, finding
$b_m = 1.9$ kpc, $b_0 = 0.1$ kpc, $\eta = 0.5$ and $\zeta = 0.6$.

In order to compute \gn\ for the bulge, I neglect the triaxiality and 
use the sphere that is geometrically equivalent to equation~\ref{bulgedist}.
This gives the same run of mass with radius, substituting
$r/[(\eta \zeta)^{1/3}b_m]$ for $b/b_m$ and similarly for $b/b_0$.  Then
\begin{equation}
g_{N,B} = \frac{V_B^2(R)}{R} = \frac{4 \pi G}{R^2}\int_0^R \rho_B(r) r^2 dr
\label{bulgevel}
\end{equation}
which I integrate numerically.  The factor $\rho_{B,0}$ is scaled to
give the correct bulge mass for each choice of scale length
(Table~\ref{MW_mods}).

The effects of deviations from this
particular bulge model are explored in \S \ref{bulgscal}.  These details
are fairly unimportant here, as they only affect the inner 3 kpc where the
motions are non-circular owing to the triaxial distribution of the inner
bulge-bar component implied by equation~\ref{bulgedist}.  
All that matters to the results at $R > 3$ kpc is the total mass enclosed therein.

\subsection{The Gas Disk \label{thegas}}

For the gas, I adopt the distribution given by Olling \& Merrifield
(2001 --- their Table D1).  I include both the molecular and atomic gas components,
treating them as being in a negligibly thin disk.  I do not include the ionized gas
component for consistency with the treatment of other galaxies.  Moreover, this
is a very small fraction of the total with an estimated surface density
($1.4\;\surfdens$) that is only available at the solar radius. 

Olling \& Merrifield (2001) give the surface densities scaled by $R_0$.
For consistency with the \THmod\ model I fix these numbers to $R_0 = 8$ kpc.
The surface densities of H$_2$ and \HI\ gas are corrected upwards by a factor
of 1.4 to account for the associated mass in helium and metals.  For these 
assumptions, the gas distribution integrates to a total mass 
$\mass_{gas} = 1.18 \times 10^{10}\;\Msun$.  This is slightly more gas mass
than inferred by Flynn \etal\ (2006) from different data, whose total sums 
to just under $10^{10}\;\Msun$.  This seems like an adequate level of agreement
considering the diversity of published opinions.

The gas is usually neglected in mass models of the Milky Way 
as it is a trace component compared to the stars and dark matter halo.  
However, the gas is not negligible in MOND.  The models considered here
have gas mass fractions in the range 
$f_g = \mass_{gas}/\mass_{b} = 0.19 \pm 0.01$
(Table~\ref{MW_mods}).  This is an important component of the 
total gravitating mass in the absence of dark matter.  Effects of the detailed
distribution of the gas are reflected in the total rotation curve.

\section{The Milky Way Rotation Curve \label{MWRC}}

Given the Milky Way mass distribution, MOND predicts the rotation curve.  
Unlike the case with external galaxies, the mass-to-light ratio is not a fit parameter.  
The \THmod\ model plus the gas distribution of Olling \& Merrifield (1999) 
specify the mass.  The only choices to be made are the scale length and the 
interpolation function.

\begin{figure}  
\epsscale{1.0}
\plotone{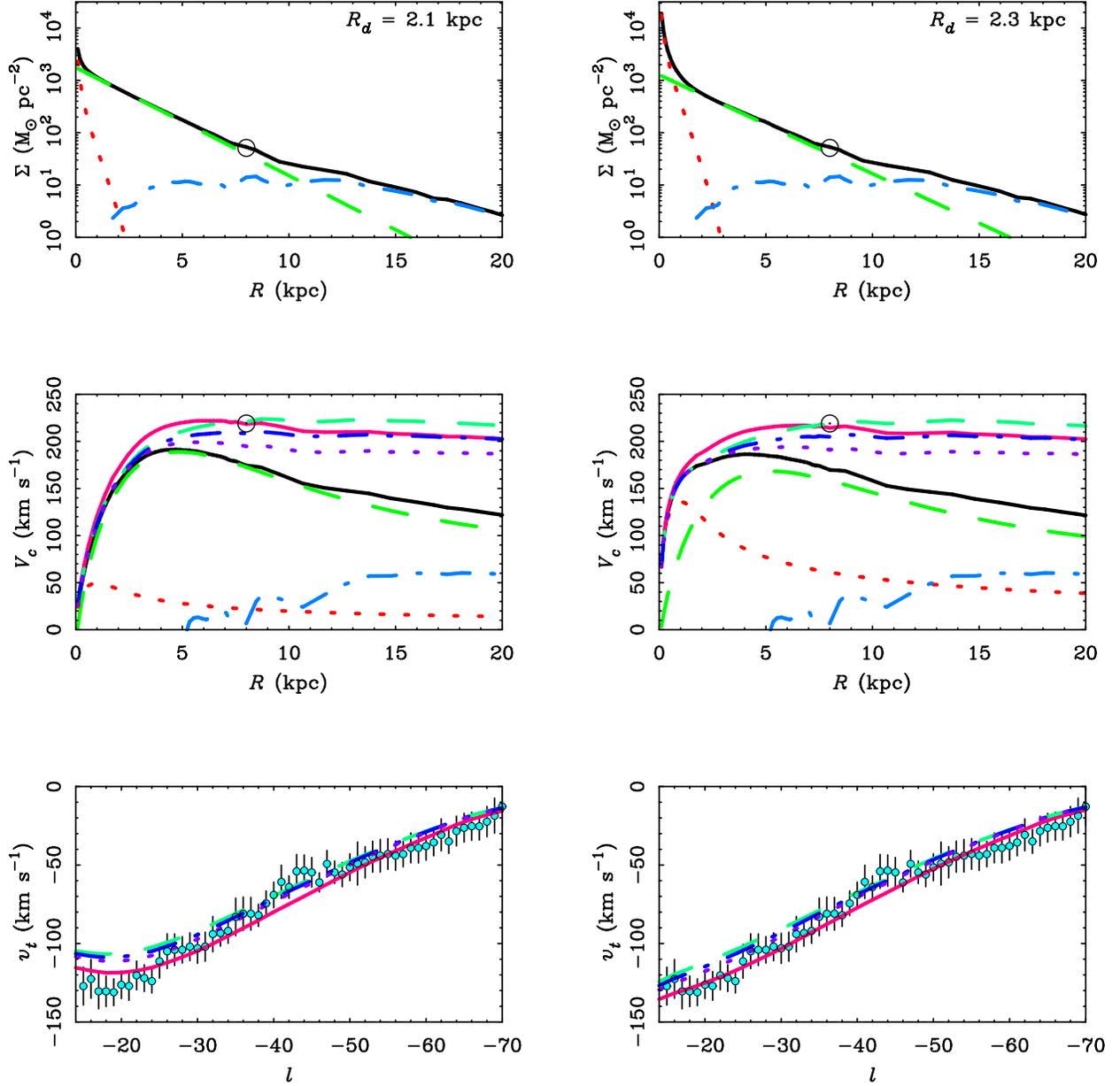}
\caption{Model Milky Ways for assumed disk scale lengths of 2.1 (left
column) and 2.3 kpc (right column).  
The top panels show the bulge (dotted line),
stellar disk (dashed line), gas (both H$_2$ and \HI; dash-dotted line), and
total (solid line) surface mass distribution.  The middle panels shows the 
Newtonian rotation curve for each of these components,
plus the rotation curve predicted by MOND for four choices of interpolation
function: $\hat \nu_1$ (solid line), $\hat \nu_2$ (dotted line), $\tilde \nu_1$
(dashed line), and $\bar \nu_1$ (dash-dotted line).  The solar value
$\Theta_0 = 219\;\kms$ at $R_0 = 8$ kpc is marked by the point.
In the bottom panels, the MOND predictions are compared to the terminal
velocity data (Kerr \etal\ 1986; Malhotra 1995).  
\label{MW_panelA}}
\end{figure}

\begin{figure}  
\epsscale{1.0}
\plotone{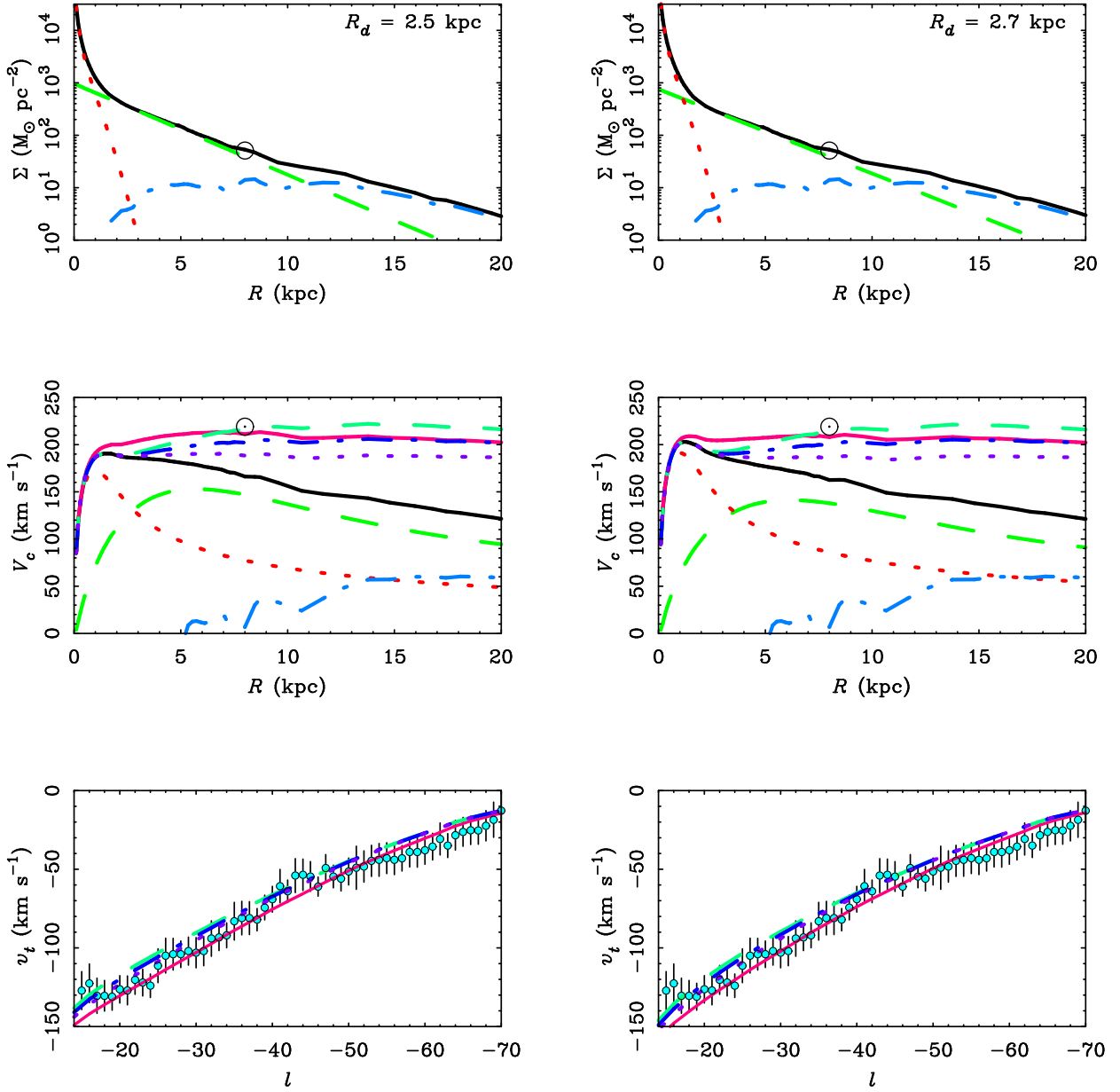}
\caption{As per Fig.~\ref{MW_panelA} but for $\rd = 2.5$ and 2.7 kpc.
\label{MW_panelB}}
\end{figure}

\begin{figure}  
\epsscale{1.0}
\plotone{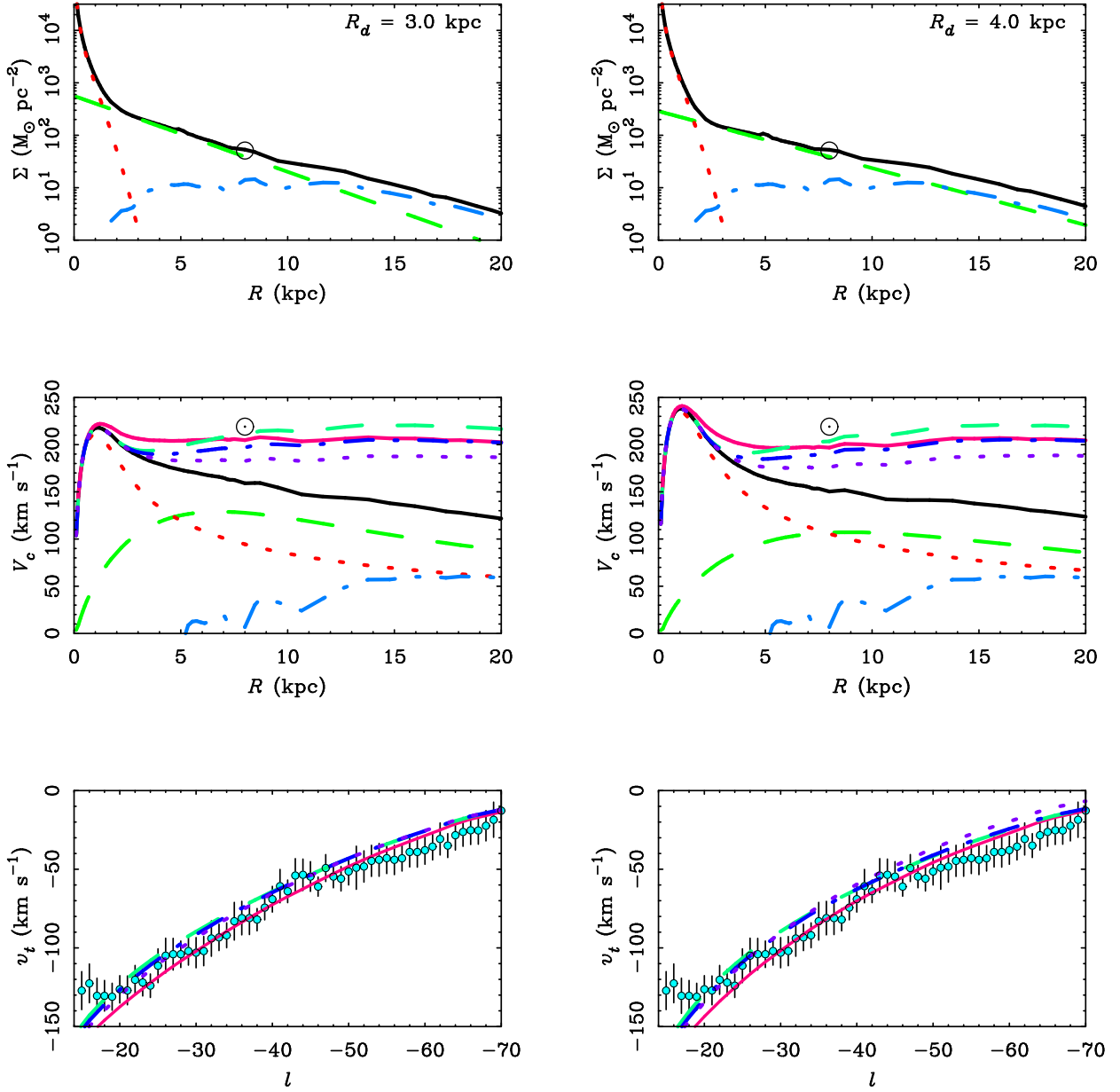}
\caption{As per Fig.~\ref{MW_panelA} but for $\rd = 3$ and 4 kpc.
\label{MW_panelC}}
\end{figure}

\placefigure{MW_panelA,MW_panelB,MW_panelC}

Figs.~\ref{MW_panelA}, \ref{MW_panelB}, and \ref{MW_panelC} 
show the results for increasing choices of scale length.  
In each case, four interpolation functions
are illustrated: $\hat \nu_1$ and $\hat \nu_2$ (these are practically 
indistinguishable from the simple and standard interpolation functions) and
for comparison the functions $\tilde \nu_1$ and $\bar \nu_1$ newly
suggested by Milgrom \& Sanders (2007).  
MOND produces a realistic rotation curve given a mass distribution,
especially for the shorter scale lengths preferred by the COBE data.

As we increase the scale length to $\rd > 3$ kpc,
MOND produces less plausible looking rotation curves.
In these cases, the bulge causes a prominent peak in the inner rotation curve.
Such a morphology is sometimes seen in early type spirals
(Noordermeer \etal\ 2007), but the sharp peak in Fig.~\ref{MW_panelC} 
is rather unusual.  This aspect is sensitive to 
the bulge model, and more plausible results are possible (\S~\ref{bulgscal}).

Comparison with the observed terminal velocities
(Kerr \etal\ 1986; Malhorta 1995) also favors short scale lengths.
The agreement is particularly good for
$\hat \nu_1$ when $\rd = 2.3$ kpc.  The other interpolation functions are
hard to distinguish from one another, and seem to prefer slightly longer
scale lengths $\rd \approx 2.5$ kpc.  This result is sensitive to small changes
(of order 4\%) in the terminal velocity data (\S \ref{MW_emp}), so stronger
statements seem unwarranted. 

The Galactic constants can be computed for each model.
Table~\ref{Oort_tab} gives the rotation velocity at the solar circle
$\Theta_0$ and the Oort Constants A and B:
\begin{mathletters}
\begin{eqnarray}
\textrm{A} = \frac{1}{2}\left(\frac{\Theta_0}{R_0} -
	\frac{d\Vc}{dR}{\Big |}_{R_0}\right) \\
\textrm{B} = -\frac{1}{2}\left(\frac{\Theta_0}{R_0} +
	\frac{d\Vc}{dR}{\Big |}_{R_0}\right).
\end{eqnarray}
\end{mathletters}
The latter depend on the derivative of the rotation curve, which in these models
depends somewhat on the extent over which the derivative is measured.
That is, there are bumps and wiggles in the rotation curve as a result of the
non-smooth gas distribution (Olling \& Merrifield 2001).  This causes the
derivative to vary in a non-trivial fashion.  For specificity, I compute A and B
over $\pm 0.5$ kpc around $R_0 = 8$ kpc.  One may wonder if this effect
has played a role in the various values of the Oort constants that have been
derived historically.

\placetable{Oort_tab}

\begin{figure}
\epsscale{1.0}
\plotone{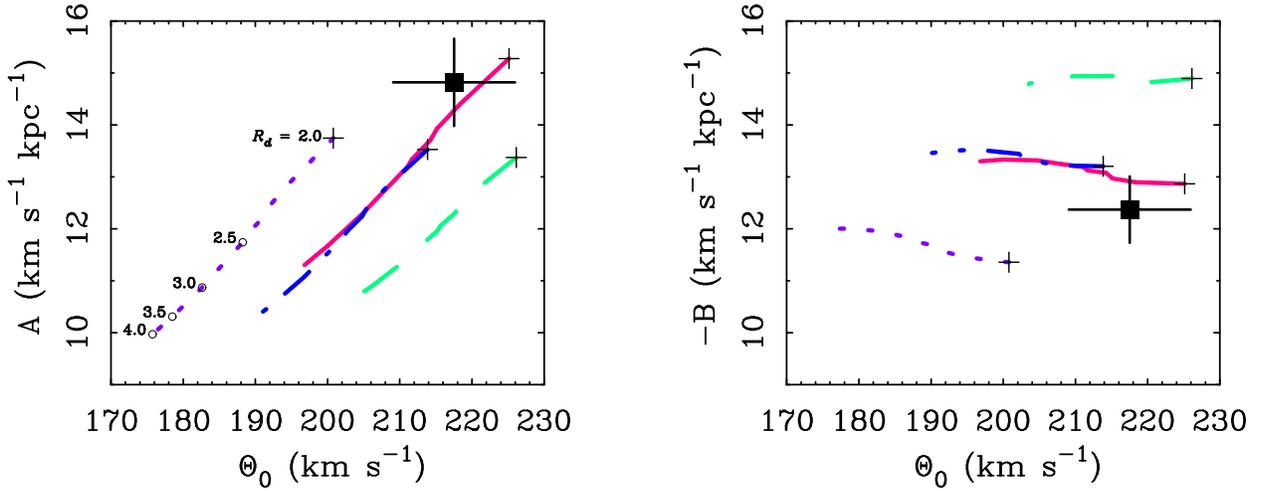}
\caption{The Oort A and B constants for the various models as a function of
$\Theta_0 = \Vc(R_0)$.  All models assume $R_0 = 8$ kpc;
see Table~\ref{Oort_tab} for the derived values of the other Galactic constants.
A cross marks the model with the shortest (2 kpc) scale length for each choice of
interpolation function (lines as per Fig.~\ref{MW_panelA}); progressively larger
\rd\ proceed along each line as illustrated by the open circles at left.  
The data points represent the observed values determined
by Feast \& Whitelock (1997).  These give 
$\Theta_0 =$ (A$-$B)$R_0 = 217.5\;\kms$, 
very close to the $219\;\kms$ obtained from the proper motion of
Sgr A$^*$ (Reid \etal\ 1999).
The precise values of A and B in the models 
depend somewhat on the range over which they are measured, 
as the bumps and wiggles 
in the gas distribution perturb the local value of the gradient in the rotation
curve (Olling \& Merrifield 2001).
\label{Oort_fig}}
\end{figure}

The Galactic constants are shown graphically in Fig.~\ref{Oort_fig}.
The measurement of Feast \& Whitelock (1997) is most consistent with
$\hat \nu_1$ for $\rd = 2.1$ kpc.  Other interpolation functions and
scale lengths are possible, depending on how literally we take the
error bars.  The function $\hat \nu_1$ seems to perform best, with reasonable
values of $\Theta_0$, A, and B for $\rd \le 2.5$ kpc.  This interpolation function
is very similar to the simple function found by Famaey \& Binney (2005)
to work best in combination with the Basel model.

The other interpolation functions perform less well, though again one must
be cautious as always about the interpretation of astronomical uncertainties.
For example, $\tilde \nu_1$ gives reasonable $\Theta_0$ up to $\rd = 3$ kpc,
but tends to give A $< |$B$|$ in contradiction to most measurements.  
Similarly, $\bar \nu_1$ gives reasonable B values,
but tends to run low in the other Galactic constants except for the smallest scale
lengths.  The standard function traditionally used in fitting external galaxies
also gives good B values for all scale lengths, but rather low $\Theta_0$,
albeit within the realm of possibility (Olling \& Merrifield 1998).

\placefigure{Oort_fig}

For all interpolation functions, agreement with both the Galactic constants and
the terminal velocities steadily deteriorates with increasing scale length.
It therefore seems clear that MOND prefers a Milky Way with a short scale
length, $\rd \lesssim 2.5$ kpc.  There also seems to be a preference
for something closer to the simple interpolation function, as found by
Famaey \& Binney (2005).
It seems quite remarkable that given the mass distribution specified by
the \THmod\ model for the stars and Olling \& Merrifield (2001) for the
gas, MOND produces a 
plausible rotation curve that is consistent with independent terminal
velocity data and the observed Galactic constants with no fitting whatsoever.

\subsection{Effects of the Bulge Scale Length  \label{bulgscal}}

The model constructed here employs the spherical radial profile that is
geometrically equivalent to the observed $L$-band light distribution
of the central bulge-bar component.  This results in a rotation curve for this
component that is very similar to that of Englmaier \& Gerhard (1999),
as it should be since it is based on the same data.  However, it
rises more steeply than that of Bissantz \etal\ (2003) owing to the
different treatment of spiral arms there.  The latter appears consistent with 
the same mass profile with a larger scale length.  

\begin{figure}  
\epsscale{1.0}
\plotone{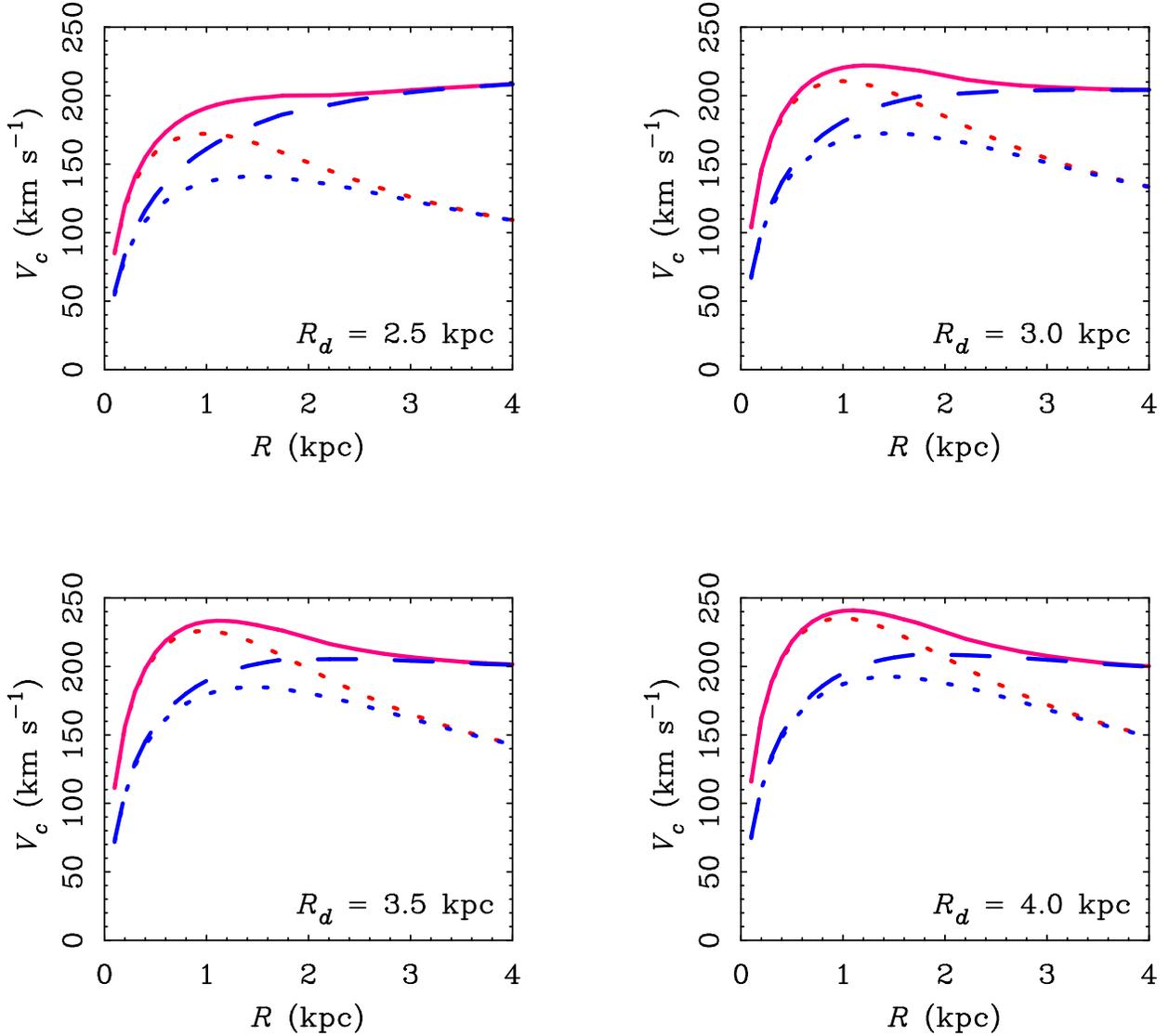}
\caption{The effect of the bulge model.  In each of the illustrated cases,
the shape of the bulge mass profile is identical.  The total mass of the bulge
changes with the scale length of the disk according to the prescription of
the \THmod\ model (Flynn \etal\ 2006).  In each panel, the two bulge models
(dotted lines) differ only in their effective scale length.  
The upper pair of lines follow from the bulge model 
computed in \S \ref{thebulge} and is similar to the model of
Englmaier \& Gerhard (1999).
The lower pair of lines makes no geometric correction
for the triaxial shape of the bulge, yet corresponds more closely to the 
model of Bissantz \etal\ (2003).  
Differences between bulge models are imperceptible outside of 3 kpc.
\label{MW_bulge_var}}
\end{figure}

Fig.~\ref{MW_bulge_var} shows the effect of varying the bulge scale length.
Two cases are illustrated:  that described in \S~\ref{thebulge}, and one with
the same distribution but a longer scale length.  The former uses the geometric
scaling $(\eta\zeta)^{1/3}$ while the latter uses no scaling.  
In effect, $r$ is directly substituted for $b$ in equation~\ref{bulgedist}
with no geometric correction to $b_m$ and $b_0$.  
This latter case gives a softer bulge more consistent with Bissantz \etal\ (2003).

\placefigure{MW_bulge_var}

The shape of the inner rotation curve is affected by the choice of scale length, 
with a much steeper rate of rise for shorter length scales.  The longer scale
length bulge model results in less of a peak, and produces morphologically
reasonable rotation curves for disks of all scale lengths.  Even the $\rd = 4$ kpc
case appears plausible, though the maximum rotation velocity barely exceeds
$200\;\kms$.  

The details of the bulge model make no difference at $R > 3$ kpc.  
Since the orbits within this radius are apparently non-circular
thanks to the triaxiality of the central component, we make no attempt to fit 
this region.  To do so would require detailed MONDian simulations that are 
well beyond the scope of this paper.  What matters here is
basically just the mass enclosed within 3 kpc, which is constrained by
a great deal of data (Gerhard 2006).  

\subsection{The Outer Rotation Curve}

\begin{figure} 
\epsscale{1.0}
\plotone{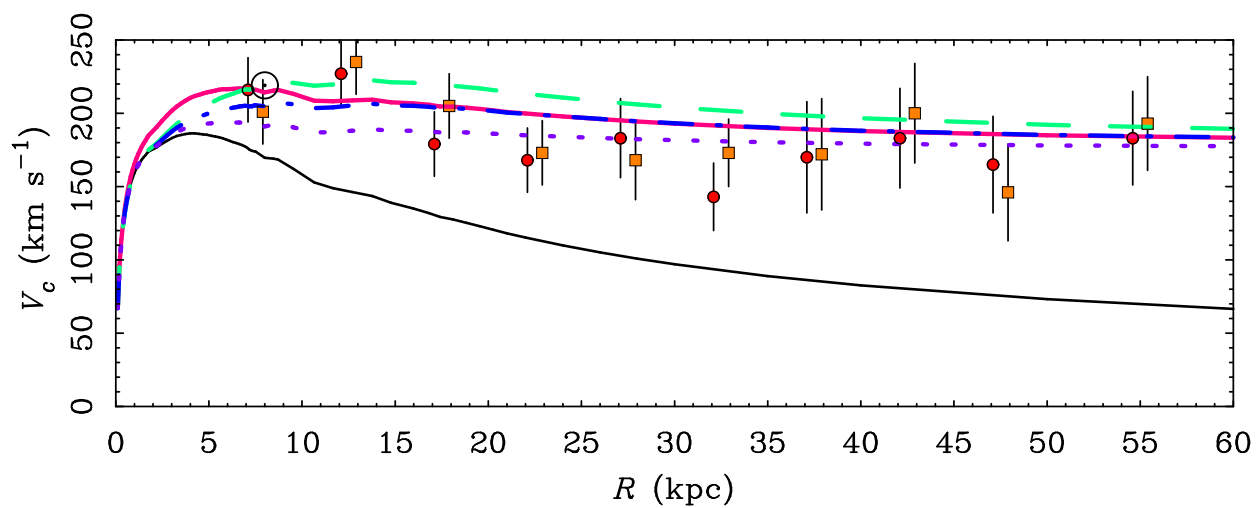}
\caption{The outer rotation curve predicted by MOND for the Milky Way 
compared to the two realizations of the Blue Horizontal
Branch stars in the SDSS data reported by Xue \etal\ (2008).
The data points from the two realizations have been offset slightly from each
other in radius for clarity;
lines as per Fig.~\ref{MW_panelA}.  The specific case illustrated has
$\rd = 2.3$ kpc, but the rotation curve beyond 15 kpc is not 
sensitive to this choice.  While the data clearly exceed the Newtonian
expectation (declining curve), they are consistent with MOND.
\label{Outer_RC}}
\end{figure}

The rotation curve of the Galaxy beyond the solar circle is rather more
difficult to constrain than that within it 
(e.g., Binney \& Dehnen 1997; Frinchaboy 2006).
MOND predicts the rotation curve for the given mass distribution quite
far out, to where the external field from other galaxies starts to become important 
(Famaey, Bruneton \& Zhao 2007; Wu \etal\ 2008).
Recently, Xue \etal\ (2008) estimated the rotation curve of the Milky Way out to
55 kpc from blue horizontal branch stars found in the Sloan Digital Sky Survey.
Fig.~\ref{Outer_RC} shows the extrapolation out to these radii.  

\placefigure{Outer_RC}

MOND was constructed to give asymptotically flat rotation curves.
However, the detailed shape of the rotation curve depends on the details
of the mass distribution.  The rotation curve may rise or fall towards the
asymptotic value, and may not become flat until quite far out.  For our Galaxy,
the rotation speed predicted by MOND declines gradually from the solar
value to $\sim 185\;\kms$ at $\sim 50$ kpc, asymptoting to $\sim 180\;\kms$.  
Such a gradually declining rotation curve helps to reconcile the apparent 
(albeit minor) discrepancy between the Milky Way and the Tully-Fisher relation 
(Flynn \etal\ 2006; Hammer \etal\ 2007): $\Theta_0$ is a bit larger than $V_f$.
More importantly, this prediction appears to be consistent with the SDSS data.

Xue \etal\ (2008) employ \LCDM\ simulations to aid in the interpretation of the
SDSS data.  This is obviously inappropriate for a MONDian analysis.  Indeed,
one wonders if it is appropriate at all given the difficulties \LCDM\ models
persistently face on galaxy scales (e.g., Kuzio de Naray \etal\ 2006;
McGaugh \etal\ 2007).  The Milky Way itself is problematic
in this regard (Binney \& Evans 2001).  
The baryon distributions of neither the \THmod\ model nor the Basel model
tolerate the expected cusp in the dark matter halo.  As with other bright spirals,
the baryons account for too much of the rotation curve budget at small radii.
Hopefully the result of Xue \etal\ (2008) is dominated by the data and 
would not change greatly with a different analysis.

\section{The Inverse Problem:  Surface Densities from Velocities \label{MW_emp}}

The \THmod\ model gives very useful constraints on the mass distribution of the
Milky Way, but is still couched in terms of a smooth exponential disk.
Real galaxies deviate somewhat from pure exponentials, and these `bumps
and wiggles' in the surface brightness are reflected in the rotation curve
(Renzo's rule:  Sancisi 2004; McGaugh 2004).  One wonders if this might also
be the case for the Milky Way.

Recently, very high quality estimates of the terminal velocities in the fourth
quadrant have become available (Luna \etal\ 2006; McClure-Griffiths \&
Dickey 2007).  These suggest the possibility of reversing the exercise above,
and inferring the surface density distribution of the Milky Way from these data.
Note that equation~\ref{nunotmu} can be inverted to obtain the baryonic
rotation curve from the observed terminal velocity curve: $\Vb = \Vc/\sqrt{\nu}$.  
Then the inversion to surface density becomes a purely Newtonian problem.
In principle, this can be accomplished by employing equation 2-174 of
Binney \& Tremaine (1987):
\begin{equation}
\Sigma(R) = \frac{1}{\pi^2 G}\left[\frac{1}{R}\int_0^R \frac{d\Vb^2}{dR}
 K\left(\frac{u}{R}\right)du + \int_R^{\infty} \frac{d\Vb^2}{dR}
 K\left(\frac{R}{u}\right)\frac{du}{u}\right].
\label{invert}
\end{equation}
Note that this procedure should work even in the context of dark matter.
If MOND is not correct as a theory, the interpolation function still provides
an empirical link between \Vb\ and \Vc\ (McGaugh 2004).
In practice, however, application of equation~\ref{invert} is fraught with peril.
The elliptic integral $K$ peaks very sharply as $u \rightarrow R$.  Moreover,
one must know the derivative of the square of the rotation curve very well.
Even though the new data are very good, there are many abrupt changes
in the derivative.  Moreover, the data extend only over a finite
range of radii, while the integral must be completed everywhere.

A more practical approach is one of iterative trial and error, computing \gn\ 
from a trial mass distribution in the usual way, then tweaking it to bring it
closer to the data.
The new terminal velocities, which are consistent in shape with the older data,
are $\sim 8\;\kms$ higher in amplitude.  This appears largely to result from the
method by which the maximum line-of-sight velocity is estimated (see
extensive discussion in McClure-Griffiths \& Dickey 2007).
While this hardly seems like a large offset ($\sim 4\%$), 
it is quite noticeable in MOND.  It implies higher surface densities, albeit
well within the uncertainties of the \THmod\ model.
Given the current inter-arm location of the sun, it might even be desirable
to have the azimuthally averaged surface density at the solar radius be somewhat
higher than the local column.

The case of $\hat \nu_1$ with $\rd = 2$ kpc is the smooth case that comes
closest to matching the data of Luna \etal\ (2006) and McClure-Griffiths \&
Dickey (2007).  Starting from this initial guess, I perturb the surface density
profile by manually adjusting the surface densities in the range necessary to
affect the terminal velocity data.  To be specific, I match the data of Luna \etal\ 
(2006) in the range $3 \leq R \leq 7.8$ kpc.  Velocities inside 3 kpc are
not fit since the motions there are thought to be non-circular,
and the details of 
the choice of bulge model begin to matter (\S \ref{bulgscal}).  For simplicity
I assume that the inner distribution is purely exponential, but this does not
matter so long as the enclosed mass remains the same.  One could just as
well imagine a Galaxy with an inner bulge plus Freeman Type II
profile that sums to the same mass.

The procedure is to adjust the surface density profile manually, estimating
the amount to adjust by the desired $\Delta \Vc^2 R$.
I then use the routine ROTMOD in GIPSY (van der Hulst \etal\ 1992)
to compute \Vb\ for the perturbed surface density distribution. 
I compare $\Vc = \hat \nu_1^{1/2}(y) \Vb$
to the data, and repeat the procedure.  Though tedious, this procedure 
can be made to converge with sufficient patience.
That is, it is possible
to obtain a model that matches the detailed shape of the terminal velocity data.

\begin{figure} 
\epsscale{1.0}
\plotone{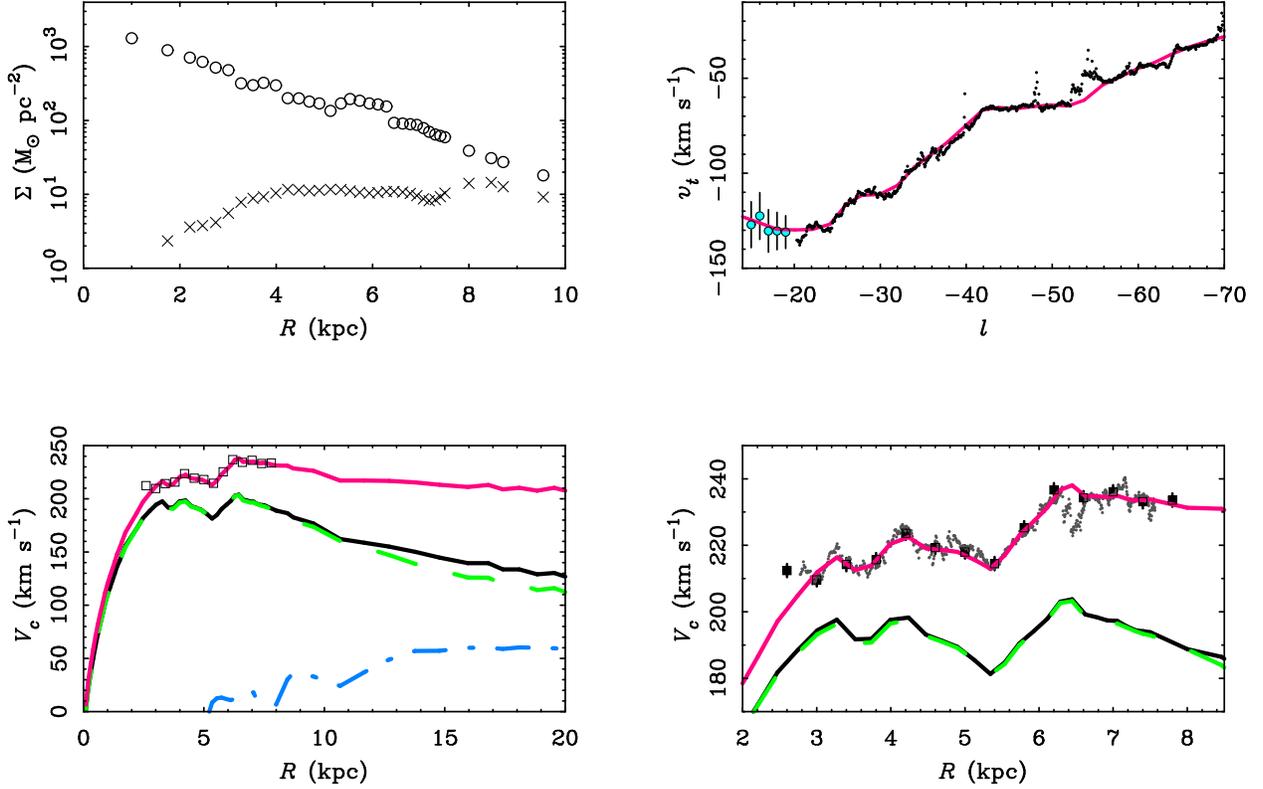}
\caption{The Milky Way stellar surface density (circles, top left) inferred from the 
terminal velocity data assuming $\hat \nu_1$ and
the illustrated gas surface densities (crosses).  
The terminal velocities of McClure-Griffiths \& Dickey (2007) 
are shown at top right together with
the MOND model fit to the data of Luna \etal\ (2006).  
The latter are shown as the square points in the lower panels.
The total and component rotation curves as per Fig.~\ref{MW_panelA} 
are shown at bottom left.  At bottom right
is a close up of the fit region with both terminal velocity data sets.
The bumps and wiggles in the velocity data can be reproduced, giving features 
similar to those seen in external galaxies in both $\Sigma(R)$ and $\Vc(R)$.  
\label{MW_empirical}}
\end{figure}

\placefigure{MW_empirical}

The result of this procedure is presented in Fig.~\ref{MW_empirical}.  
Note that in order to affect the velocity at 3 kpc, it is necessary to start adjusting 
the surface density somewhere inside of that.  The inferred stellar surface 
densities and corresponding velocities are given in Table~\ref{MWSD_emp}.
Outside of this range, the stellar density remains that of a purely
exponential disk.  The gas is assumed to follow the distribution of
Olling \& Merrifield (2001); only the stellar disk has been adjusted.
The total mass is 2\% higher than the initial pure exponential disk:
$\mass_{disk} = 5.48 \times 10^{10}\;\Msun$.

\placetable{MW_empirical}

We should be careful not to over-interpret the result.  I have only fit one choice of
interpolation function ($\hat \nu_1$); other choices would give somewhat different
results.  Moreover, the assumption of circular motion is implicit; streaming motions
along the spiral arms are likely to be present at some level. 
Indeed, Luna \etal\ (2006) give $\Vc(R=7.8\;\textrm{kpc}) = 233.6\;\kms$.  
This is difficult to reconcile with $\Theta_0(R_0=8.0\;\textrm{kpc}) = 219\;\kms$ 
(Reid \etal\ 1999) with purely circular motion in any type of model.
Variations of this sort are at least conceivable in MOND 
(Fig.~\ref{MW_empirical}), but probably reflect a real difference between the
first and fourth quadrants.  Hence I have made no attempt to force a fit to
the solar value.  

The Oort constants of this model are fairly reasonable:
$\textrm{A} = 15.9\;\kms\,\textrm{kpc}^{-1}$ and
$\textrm{B} = -13.0\;\kms\,\textrm{kpc}^{-1}$.  The value of A may seem
a bit high, but note that since the rotation velocity is inferred to be higher than 
the solar value, A$-$B must also be higher.  This is in the data.
The model fits the detailed terminal velocity curve as far as it is 
reported (up to $R = 7.8$ kpc), so these are in effect the measured values of the 
Oort constants in the fourth quadrant.  There is only a modest model dependent
extrapolation to the solar radius.
Barring systematic errors in the data or sharp features in the
rotation curve near the solar radius, the uncertainty in these
estimates is $< 1\;\kms\,\textrm{kpc}^{-1}$.

Indeed, it is instructive that this exercise can be successfully done at all.
The inferred surface density has the sorts of bumps and wiggles commonly
observed in the azimuthally averaged surface brightness profiles of spiral galaxies.
These correspond to the bumps and wiggles in the rotation curve, as they must
in MOND, and as they are observed to do in general (Renzo's rule).  
This correspondence follows
in the dark matter picture only if disks are dynamically important.  This is hard
to arrange with the cuspy halos obtained in CDM simulations (e.g., 
Navarro, Frenk, \& White 1997) as these place too much dark mass at
small radii.  Low surface
brightness disks can not have dynamically significant mass in the dark matter
picture (de Blok \& McGaugh 1997), yet still obey the correspondence of bumps
and wiggles encapsulated by Renzo's rule (e.g., Broeils 1992).  This occurs
naturally in MOND, the \textit{a priori} predictions of which (Milgrom 1983b) are
realized in LSB galaxies (Milgrom \& Braun 1988; McGaugh \& de Blok 1998).

The specific pattern of bumps and wiggles seen in Fig.~\ref{MW_empirical}
is in principle testable by star count analyses.  In particular, it is tempting to
associate the dip in surface density at $\sim 5$ kpc and the subsequent shelf
with a ring or spiral arms, perhaps emanating from the end of the long 
($\sim 4.5$ kpc) bar (Cabrera-Lavers \etal\ 2007) --- 
a morphology frequently seen in other
galaxies and naturally reproduced in MOND simulations (Tiret \& Combes 2007).  
Again however, the details of star counts will depend on the choice of interpolation
function and the level of non-circular motions.  Indeed, even for a given
interpolation function and purely circular motion, the result at this level of
detail depends on whether we use the modified gravity of Bekenstein \& Milgrom
(1984; equation~\ref{AQUAL}) or modified inertia (equaion~\ref{mondeqn}).
I have implicitly assumed the latter here.

Another intriguing thing to note is that a fit to the surface densities in 
Table~\ref{MWSD_emp} gives $\rd = 2.4$ kpc even though the
base model has $\rd = 2.0$ kpc.  This type of variation in \rd\ with the fitted 
radial range is commonly found in external galaxies, and may go some way 
to explaining the variation in reported scale lengths for the Milky Way
(Sackett 1997).  Clearly MOND prefers a compact Milky Way, consistent 
with the COBE data (Binney \etal\ 1997; Drimmel \& Spergel 2001) and the
rather high observed microlensing optical depth (Popowski \etal\ 2005)
that requires that baryonic mass dominate within the solar circle 
(Binney \& Evans 2001; Bissantz, Debattista, \& Gerhard 2004).

A related test of MOND in the Milky Way is provided by the vertical support
of the stellar and gas disks.  In the outskirts of the galaxy, the baryonic components
should flare substantially as the effective mass scale height comes to be dominated
by the quasi-spherical dark matter halo.  In contrast, the potential remains disk-like
close to the plane in MOND.  The net effect is that MONDian disks will be
somewhat thinner, all other things being equal.

Recently, S{\'a}nchez-Salcedo, Saha, \& Narayan (2008) have analyzed the thickness 
of the gas layer of the Milky Way in the context of MOND.  They find that
MONDian self-gravity of the disk provides a plausible explanation of the
thickness and flaring of the gas layer.  Milgrom (2008) points out that the
form of the interpolation function as well as the mass model matters to the
details of the support of the gas layer.  It also seems possible that 
magnetic field support may be non-negligible, so a perfect fit might be
difficult to obtain.  In comparison, it is necessary to invoke a massive 
($> 10^{11}\;\Msun$) ring-like dark matter
component in addition to the traditional quasi-spherical halo in order to 
explain the gas thickness in conventional terms (Kalberla \etal\ 2007).

For stars, MOND will support a higher vertical velocity dispersion at a
given disk thickness.  This effect in the Milky Way is rather subtle 
until large radii (Nipoti \etal\ 2007), but might conceivably be detectable 
with GAIA.  The large velocity dispersions of planetary nebulae at large 
radii in face-on spirals might be an indication of such an effect 
(Herrmann \& Ciardullo 2005).

\section{Conclusions \label{conc}}

Treating the Milky Way as we would an external galaxy, it is possible to
obtain the rotation curve from the surface density with MOND.
There is no freedom to adjust the mass-to-light ratio as in external galaxies.
The result is satisfactory provided the scale length of the Milky Way is
relatively short (2 --- 2.5 kpc), as implied by the COBE data (Binney \etal\ 1997;
Gerhard 2002, 2006).  It is also possible to invert the procedure and derive a
plausible detailed surface density distribution from the observed terminal
velocity curve.

The major conclusions of this work are as follows.
\begin{itemize}
\item Given the observed stellar and gas mass distribution of the Milky Way,
MOND naturally produces a plausible rotation curve that is consistent with
the relevant dynamical data.  This follows with no fitting.

\item MOND prefers Milky Way models with relatively short 
($2.0 \lesssim \rd \lesssim 2.5$ kpc) scale lengths.  In this range, rotation
curves that look familiar from the study of external galaxies are produced.
As the scale length is increased beyond this range,
the morphology of the resulting rotation curve becomes less realistic,
and the match to kinematic data becomes worse.

\item The Milky Way data seem to prefer an interpolation function close to
the simple form, in agreement with the findings of Famaey \& Binney (2005).
The precise form that is preferred depends on the details of the adopted 
Milky Way model, so it is unclear how definitive a statement can be made.

\item An interpolation function that shares the virtues of
the simple function on galaxy scales without having as large an impact on
solar system dynamics is $\hat \nu_1^{-1}(y) = 1 - e^{-\sqrt{y}}$.  Other forms
are possible.  Empirical calibration of this function is desirable, even in
the context of dark matter, since it encapsulates the coupling between mass
and light.

\item It is possible to recover the detailed surface mass density of the Milky Way
from the observed terminal velocities.  The result is a Galaxy
with bumps and wiggles in both its luminosity profile and rotation curve
that are reminiscent of those frequently observed in external galaxies.

\item A prominent feature among the bumps and wiggles is a shelf around
$\sim 5.5$ kpc.  This might correspond to a ring or spiral arms, perhaps 
extending from the ends of the long bar.  Fitting the profile including the
bumps and wiggles gives $\rd = 2.4$ kpc even though the mass scale length
is 2 kpc.

\item The Oort constants in the fourth quadrant are estimated to be 
$\textrm{A} = 15.9\;\kms\,\textrm{kpc}^{-1}$ and
$\textrm{B} = -13.0\;\kms\,\textrm{kpc}^{-1}$.

\end{itemize}
It is hard to imagine that all this could follow from a formula devoid of
physical meaning.

\acknowledgements 
The author is grateful for conversations with Moti Milgrom, Garry Angus, and
Benoit Famaey, and for constructive input from the referee.  
The work of SSM is supported in part by NSF grant AST 0505956.

\begin{deluxetable}{ccccc}
\tablewidth{0pt}	
\tablecaption{Milky Way Models\label{MW_mods}}	
\tablehead{
  \colhead{\rd} & \colhead{\Sd} & \colhead{$\mass_{disk}$}
   & \colhead{$\mass_{bulge}$} & \colhead{$\mass_{b}$} \\
  \colhead{(kpc)} & \colhead{($\surfdens$)} &
  \multicolumn{3}{c}{($10^{10}\;\Msun$)} 
}
       \startdata	
2.0 &	2133 &	5.36 &	0\phd\phn\phn & 6.54	 \\
2.1 &	1765 &	4.89 &	0.09 & 6.16	 \\
2.2 &	1480 &	4.50 &	0.40 & 6.08	\\
2.3 &	1270 &	4.22 &	0.67 & 6.07	 \\
2.4 &	1097 &	3.97 &	0.91 & 6.06	 \\
2.5 &	\phn 960 &	3.77 &	1.10 & 6.05	 \\
2.7 &	\phn 755 &	3.46 &	1.37 & 6.01	 \\
3.0 &	\phn 562 &	3.18 &	1.66 & 6.02	 \\
3.5 &	\phn 383 &	2.95 &	1.91 & 6.04	 \\
4.0 &	\phn 287 &	2.89 &	2.07 & 6.14	 \\
\enddata	
\tablecomments{The central surface density \Sd\ is inferred from
the scale length-disk mass-bulge mass relation of the \THmod\ 
model (Flynn \etal\ 2006).  The total baryonic mass 
$\mass_{b} = \mass_{disk}+\mass_{bulge}+\mass_{gas}$ 
is the sum of all known non-negligible baryonic components.}
\end{deluxetable}	
	
\clearpage

\begin{deluxetable}{ccccccccccccc}
\tablewidth{0pt}	
\tablecaption{Galactic Constants\label{Oort_tab}}	
\tablehead{
\colhead{} & \multicolumn{3}{c}{$\hat \nu_1$} &
  \multicolumn{3}{c}{$\tilde \nu_1$} & \multicolumn{3}{c}{$\bar \nu_1$} &
  \multicolumn{3}{c}{$\hat \nu_2$} \\
  \colhead{\rd} & 
  \colhead{$\Theta_0$} & \colhead{A} & \colhead{B} &
  \colhead{$\Theta_0$} & \colhead{A} & \colhead{B} &
  \colhead{$\Theta_0$} & \colhead{A} & \colhead{B} &
  \colhead{$\Theta_0$} & \colhead{A} & \colhead{B} 
}
       \startdata	
2.0 & 225 &     15.2 &  $-$12.8 &       226 &   13.3 &  $-$14.8 &       214 &   
13.5 &  $-$13.2 &       201 &    13.7 &  $-$11.3 \\
2.1 & 218 &     14.3 &  $-$12.8 &       221 &   12.7 &  $-$14.8 &       208 &   
12.8 &  $-$13.2 &       195 &    12.8 &  $-$11.4 \\
2.2 & 215 &     13.9 &  $-$12.9 &       218 &   12.4 &  $-$14.8 &       206 &   
12.4 &  $-$13.2 &       192 &    12.4 &  $-$11.5 \\
2.3 & 214 &     13.7 &  $-$13.0 &       218 &   12.3 &  $-$14.9 &       205 &   
12.2 &  $-$13.3 &       191 &    12.2 &  $-$11.6 \\
2.4 & 212 &     13.3 &  $-$13.1 &       216 &   12.0 &  $-$14.9 &       203 &   
11.9 &  $-$13.3 &       189 &    11.8 &  $-$11.7 \\
2.5 & 211 &     13.1 &  $-$13.1 &       215 &   11.9 &  $-$14.9 &       202 &   
11.8 &  $-$13.4 &       188 &    11.7 &  $-$11.7 \\
2.7 & 208 &     12.7 &  $-$13.2 &       213 &   11.6 &  $-$14.9 &       199 &   
11.4 &  $-$13.4 &       185 &    11.3 &  $-$11.8 \\
3.0 & 205 &     12.2 &  $-$13.3 &       210 &   11.3 &  $-$14.9 &       197 &   
11.0 &  $-$13.5 &       183 &    10.8 &  $-$11.9 \\
3.5 & 200 &     11.6 &  $-$13.3 &       206 &   10.8 &  $-$14.8 &       193 &   
10.5 &  $-$13.5 &       179 &    10.3 &  $-$12.0 \\
4.0 & 197 &     11.3 &  $-$13.2 &       204 &   10.6 &  $-$14.7 &       190 &   
10.3 &  $-$13.4 &       176 &    10.0 &  $-$11.9 \\
\enddata	
\tablecomments{Galactic constants as predicted by each interpolation function
$\nu$ for each choice of disk scale length.  Galactic units are used:  \rd\ is in kpc,
$\Theta_0$ is in $\kms$, and A and B are in $\kms\;\textrm{kpc}^{-1}$.}
\end{deluxetable}	
	
\clearpage

\begin{deluxetable}{ccc}	
\tabletypesize{\footnotesize}
\tablewidth{0pt}	
\tablecaption{Inferred Surface Densities\label{MWSD_emp}}	
\tablehead{
  \colhead{$R$} & \colhead{\Sst} & \colhead{\Vc} \\
  \colhead{(kpc)} & \colhead{($\surfdens$)} & \colhead{($\kms$)} }
       \startdata	
       2.47    &   620 &	 197 \\
       2.74    &   520 &	 204 \\
       3.01    &   480 &	 211 \\
       3.27    &   318 &	 216 \\
       3.52    &   300 &	 212 \\
       3.74    &   323 &	 213 \\
       3.99    &   299 &	 220 \\
       4.24    &   200 &	 222 \\
       4.47    &   199 &	 219 \\
       4.69    &   180 &	 218 \\
       4.90    &   170 &	 218 \\
       5.18    &   135 &	 215 \\
       5.34    &   170 &	 213 \\
       5.54    &   195 &	 216 \\
       5.74    &   185 &	 223 \\
       5.94    &   170 &	 227 \\
       6.11    &   165 &	 231 \\
       6.29    &   155 &	 236 \\
       6.45    &   \phn 93 &	 238 \\
       6.62    &   \phn 91 &	 234 \\
       6.78    &   \phn 89 &	 234 \\
       6.92    &   \phn 87 &	 234 \\
       7.06    &   \phn 79 &	 234 \\
       7.18    &   \phn 70 &	 234 \\
       7.30    &   \phn 65 &	 233 \\
       7.41    &   \phn 62 &	 233 \\
       7.50    &   \phn 59 &	 233 \\
\enddata	
\tablecomments{Stellar surface densities outside this range follow an 
exponential distribution with $\rd = 2.0$ kpc and $\Sd = 2133\;\surfdens$.}
\end{deluxetable}	

\clearpage

\end{document}